\documentstyle[11pt]{article}
\begin{document}
\thispagestyle{empty}

\begin{center}

RUSSIAN GRAVITATIONAL ASSOCIATION\\
CENTER FOR SURFACE AND VACUUM RESEARCH\\
DEPARTMENT OF FUNDAMENTAL INTERACTIONS AND METROLOGY\\
\end{center}
\vskip 4ex
\begin{flushright}                              RGA-CSVR-015/94\\
                                                gr-qc/9412003
\end{flushright}
\vskip 45mm

\begin{center}
{\bf Quantum creation of qusihomogeneous inflationary universe}

\vskip 5mm
{\bf A. A. Kirillov}\\

\vskip 5mm
{\em Institute for
Applied Mathematics and Cybernetics, 10 Uljanova Str.}\\
{\em Nizhny Novgorod, 603005, Russia}\\
e-mail:kirillov@focus.nnov.su

\vskip 5mm

\begin{abstract}
The process of quantum creation of a qusihomogeneous inflationary universe
near a cosmological singularity is considered. It is shown that during the
evolution quantum fluctuations of spatial topologies increase and the
universe acquires homogeneous and isotropic character on arbitrary large
distances.\end{abstract}

\vskip 30mm

           Moscow 1994
\end{center}
\pagebreak

\setcounter{page}{1}
As is well known a natural explanation of the initial conditions for standard
cosmological models can be given in the framework of inflationary universe
scenarios \cite{inf,spect}. Nevertheless, it was pointed out \cite{alstar83}
that on sufficiently large distances exceeding the visible part of the
present universe one could expect the universe to be essentially
inhomogeneous and anisotropic. In this paper we show that it will not be so
if quantum topology fluctuations (the spacetime foam
\cite{wheeler,hawking78}) are taken into account. More precisely, we show
that during the evolution topology fluctuations increase and if in the very
beginning the universe had a rather simple topology the spacetime foam will
almost completely determine properties of the universe.

The simplest processes connected with topology changing (wormholes and baby
universes) are known to be described in the framework of third quantization
\cite{wormholes}. In particular, third quantization is the natural tool for
description of quantum creation of a universe from nothing
\cite{qntcreat,kir92}.  In present paper we use a new approach pointed
out in Ref.\cite{kir94} which generalizes the third quantization and allows
to describe arbitrary topologies of the universe.  That generalization
follows from the fact that the Wheeler-DeWitt equation consists of an
infinite set of Klein-Gordon type equations (there is one local
Wheeler-DeWitt equation at each point of a coordinate basic manifold $x\in
S$). Moreover, in a number of cases (close to a cosmological singularity;
inflationary stage in the evolution of the universe) these local equations
become in a leading order uncoupled, since every local WDW equation contains
variables (metric functions and scalar fields) specified at the given point
$x$ of the coordinate manifold $S$. If we assume now that the number of
physical points corresponding to an arbitrary point of the basic manifold $S$
may be a variable we come to the pointed generalization for third
quantization.

Let us consider an inhomogeneous universe filled with a number of scalar
fields $\phi ^a$ $a=1,...,d$. Using Kasner-like variables \cite{kir93} the
action can be represented in the following form (we use Plankian units
$l_{pl}=1$ and suppose $\partial S=0$) \begin{equation} \label{eq:actinfl}
I=\int_{S}[p_A {\partial \over \partial t} z^A - \lambda (p_i^2 -p_0^2
+U(z,p)] d^{3}x dt \end{equation} where ($A=0,...,d+2$), $\lambda$ is
expressed via the lapse function and $z^A$ is a set of scalar fields and
logarithms of metric scale functions \cite{kir93} (3-metric takes the form
$g_{\alpha \beta}(x)=\sum_{n,m=0}^2 \exp (A^n_m z^m
(x))l^n_{\alpha}(x)l^n_{\beta}(x)$ with constant matrix $A^n_m$).  The
potential term in (\ref{eq:actinfl}) is $U=6g(W- ^3 R)$ (here $^3 R$ is the
scalar curvature and $W=W(\phi ,\partial _{\alpha }\phi, g_{\alpha \beta})$
is given by a potential for scalar fields) and due to solving the momentum
constraints $U$ is a function of all dynamical variables $p_A (x)$ and $z^A
(x)$. The inflationary stage in the evolution of the universe begins under
the following conditions \begin{equation} \label{eq:infcond} ^3R\ll W, \qquad
W\approx \Lambda= const, \end{equation} which imply the potential becomes an
effective cosmological constant \cite{inf,alstar83}. The conditions
(\ref{eq:infcond}) imply also defined restrictions on the degree of
inhomogeneity of the universe.  In quantum gravity in order to get
consistency of the model we shall also assume the existence of a
sufficiently small minimal scale $l_{min}$ for inhomogeneities and thereby
considering short-distance fluctuations to be omitted. For the sake of
simplicity in what follows we put $l_{min}=1$.

The configuration space $M$ of the system (\ref{eq:actinfl}) (called also
superspace) can be represented in the form of the direct product
$M=\prod_{x\in S}M_x$, where $M_x$ is the ordinary $d+3$-dimensional
pseudo-Euclidian space. Quantization of the system (\ref{eq:actinfl}) leads
to the Wheeler-DeWitt equation ($g= e^{3z^0}$)
\begin{equation}\label{eq:wdwinfl} (\Delta _{x}+ 6\Lambda e^{3z^0(x)}) \Psi =
0 , \qquad x\in S, \end{equation} where $\Delta _{x}$ denotes a Laplace
operator on $M_x$ and $\Psi$ is a wave function of the universe.  Since the
equations (\ref{eq:wdwinfl}) turn out to be uncoupled the space $H$ of
solutions to these equations has the form of the tensor product of spaces
$H_{x}$ ($H =\prod_{x\in S}H_{x})$ as that of $M$, where $H_{x}$ is the space
of solutions to a separate $x-$ equation (\ref{eq:wdwinfl}).  Then we can
introduce a local wave function $\Psi _x$ specified on $W_x$ and describing
quantum states of the scalar fields and three-geometry in a neighbourhood of
a particular point $x\in S$.

Let us now assume that the number of points of the observable physical space
may be a variable. This means that at a particular supporting point of the
coordinate manifold $x\in S$ there is a number of points corresponding to the
physical space. In quantum theory this fact can be accounted by third
quantization of the every local wave function $\Psi_x$ introduced above. The
last ones become field operators and can be expanded in the form (for
simplicity we consider $\Psi _x$ to be a real scalar function)
\begin{equation}\label{eq:wavef} \Psi _x  = \sum
C(n,x)U(n,x)+C^{+}(n,x)U^{*}(n,x), \end{equation} where
$\{U(n,x),U^{*}(n,x)\}$ is an arbitrary complete basis in $H_x$ and the
operators $C(n,x)$ and $C^{+}(n,x)$ satisfy the standard commutation
relations \begin{equation}\label{eq:commrel} [C(n,x),C^{+}(m,y)]=\delta
_{n,m}\delta (x,y).  \end{equation} The field operators $\Psi_x$ act on a
Hilbert space of states which has well known structure in Fock
representation.  The vacuum state is defined by the relations $C(x,n)\mid
0>=0$ (for all $x\in S$), $<0|0>=1$.  Acting by the creation operators
$C^{+}(n,x)$ on the vacuum state we can construct states describing a
universe with arbitrary spatial topologies.  In particular, the states
describing the ordinary universe have the structure
\begin{equation}\label{eq:consts} |f> =\sum_{[n(x)]} F_{n(x)}
|1_{n(x)}> , \qquad |1_{n(x)}>={1\over Z_1}\prod_{x\in S}C^{+}(x,n(x))|0>,
\end{equation} where $Z$ is a normalization constant.
The wave function describing a simple universe takes the form
\begin{equation} \label{eq:simplwf}
<0|\Psi |f> =<0|\prod_{x\in S}\Psi_x |f>=\sum_{[n(x)]} F_{n(x)}u[n(x)]
\end{equation}
where $u[n(x)]=\prod_{x\in S}U(n(x),x)$.  The states describing a universe
with $n$ disconnected spatial components has the following structure
\begin{equation}\label{eq:tqsts} |n>=|1_{m_1 (x)},...,1_{m_n (x)}>={1\over
Z_n}\prod_{i=1}^n\prod_{x\in S}C^{+}(x,m_i (x))|0> \end{equation} (we remind
that in the model under consideration due to existence of $l_{min}$ the
coordimates $x$ take discrete values).  Besides these states describing
simplest topologies the considered approach allows to construct nontrivial
topologies as well.  This is due to the fact that the tensor product in
(\ref{eq:consts}), (\ref{eq:tqsts}) may be defined either over the whole
coordinate manifold $S$ or over a part of it $K\subset S$. In this manner,
taking sufficiently small pieces $K_i$ of the coordinate manifold $S$ we can
glue arbitrarily complex physical spaces. In order to construct the states of
such a kind it turns out to be convenient to introduce the following set of
operators \begin{equation} \label{eq:gcreat} a(K,n(K))=\prod_{x\in
K}C(x,n(x)),\qquad a^{+}(K,n(K))=\prod_{x\in K}C^{+}(x,n(x)).  \end{equation}
These operators have a clear interpretation, e.g. the operator
$a^+(K,n(K))$ creates the whole region $K\in S$ having the quantum numbers
$n(K)$. Thus, in the general case  states of the universe will be described
by vectors of the type \begin{equation} \label{eq:gsts} |\Phi>=c_0|0>+\sum_I
c_I a^+_I|0> +\sum_{I,J}c_{IJ}a^+_I a^+_J|0>+... .  \end{equation}

Now consider the interpretation of the suggested approach.
Ordinary measurements are usually performed only on a part $K$ of the
coordinate manifold $S$. There are two possibilities. The first one is when
an observer measures all of the quantum state of the region $K$ and the
second more probable one is when the observer measures only a part of the
state. In the second case the observer considers $K$ as if it were a part of
the ordinary flat space. Therefore, the part of the quantum state which will
be measured, appears to be in a mixed state. This means the loss of quantum
coherence widely discussed in Refs.\cite{wormholes}.  In order to describe
measurements of the second type we define the following
density matrix for the region $K$
\begin{equation} \label{eq:denmatrix} \rho ^{nm}(K)={1\over N(K)}<\Phi
|a^{+}(K,n(K)) a(K,m(K))|\Phi >, \end{equation} where $|\Phi>$ is an
arbitrary state vector of the (\ref{eq:gsts}) type and $N(K)$ is a
normalization function which measures the difference of the real spatial
topology from that of the coordinate manifold $S$.  For the states
(\ref{eq:consts}) we have $N(K)=1$.
Thus, if $A(K)$ is any observable we find $<A>={1\over N}Tr(A\rho )$.

If we consider the smallest region $K$ which contains only one point $x$ of
the space $S$ the normalization function $N(x)$ in (\ref{eq:denmatrix}) will
play the role of a "density" of the physical space and the states
(\ref{eq:consts}), (\ref{eq:tqsts}) give $N(x)=1$ and $N(x)=n$ respectively.

The distinctive feature of the WDW Eq.(\ref{eq:wdwinfl}) is the fact that it
has explicit "time"-dependent form. Therefore, one could expect the existence
of quantum polarization effects (topology fluctuation or the so-called
spacetime foam
\cite{wheeler,hawking78}).  These effects can be calculated either by singling
out the asymptotic {\em in} and {\em out} regions on the configuration space
$M$ for which we can determine positive-frequency solutions to
Eq.(\ref{eq:wdwinfl}) (see for example \cite{kir92}), or by using the
dioganalization of Hamiltonian technique
\cite{grmm} by means of calculating depending on time Bogoliubov's
coefficients.

Let us consider solutions to an arbitrary local $x$-equation
(\ref{eq:wdwinfl}).  These solutions can be represented in the form $u(p,x)=
(2\pi)^{-{d+2\over 2}}e^{ipz}\varphi_p (z^{0})$ where $\varphi$ satisfies the
equation
\begin{equation}\label{eq:InfWdw} {d^2 \varphi _p \over d^2 z^{0}} +
\omega^2_p (z^0 )\varphi _p =0, \qquad \omega^2 = p^2 +6\Lambda e^{3z^0}
\end{equation} and is expressed in terms of Bessel or Hankel functions.  The
function $\varphi $ can be decomposed in positive and negative frequency
parts
\begin{equation} \label{eq:sols} \varphi _p ={1\over
\sqrt {2\omega_p}}(\alpha _p e^{i\theta _p} +\beta _p e^{-i\theta _p}), \qquad
{d
\varphi _p \over d z^{0}}= i \sqrt {{\omega_p\over 2}}(\alpha _p e^{i\theta
_p} -\beta _p e^{-i\theta _p}), \end{equation} where $\theta _p =\int^{z^0}
\omega _p dz^0$ .  The functions $\alpha _p$ and $\beta _p$ satisfy identity
$|\alpha _p|^2-|\beta_p|^2=1$ and define the depending on time Bogoliubov
coefficients \cite{grmm}.  Now we determine two asymptotic regions as {\em
in} ($z^0\rightarrow -\infty$) and {\em out} ($z^0\rightarrow +\infty$). In
these regions the functions $\alpha _p$ and $\beta _p$ take constant values
and therefore, in these regions we can define positive frequency functions as
$U(p,x)= (2\omega_p(2\pi)^{d+2})^{-{1\over 2}}e^{ipz+i\theta _p}$.
Substituting
the initial conditions $\alpha _p =1$, $\beta _p=0$ as
$z^0\rightarrow -\infty$ in (\ref{eq:InfWdw}), (\ref{eq:sols}) we find
that in the {\em out} region the Bogoliubov
coefficients are \begin{equation}\label{eq:bgcoef} \alpha_p =(\exp({3\pi
p\over 2})/ 2sh({3\pi p\over 2}))^{{1\over 2}} ,\qquad \beta_p =(\exp(-{3\pi
p\over 2})/ 2sh({3\pi p\over 2}))^{{1\over 2}}. \end{equation}
Then, for example, if the initial  state of the "superspace"-Hamiltonian is the
ground state $|0_{in}>$ in the {\em out} region the
density matrix
(\ref{eq:denmatrix}) takes form
\begin{equation} \label{eq:infden}
\rho ^{pq}(K)=\prod _{x\in K} \rho ^{p(x)q(x)}(x),
\end{equation}
where $\rho (x)$ is a
one-point density matrix
\begin{equation} \label{eq:xmatrix}
\rho ^{pq}(x)={1\over N(x)}|\beta_p|^2\delta (p,q)=
{1\over N(x)}{1\over e^{3\pi p} -1}\delta (p,q) .
\end{equation}
The normalization function in (\ref{eq:xmatrix}) is given by $N(x)=V_x n_d$
(where $V_x$ is the spatial volume of the configuration space $M_x$ and
$n_d$ is a constant).  The matrix (\ref{eq:infden}) does not depend on
spatial coordinate and has the Plankian form with the temperature $T={1\over
3\pi }$
and therefore, we obtain the
creation of a universe which in average turns out to be homogeneous.

In conclusion we note that the property of the created universe to be
homogeneous
follows in the first place from the specific choice of
the homogeneous initial quantum state $|0_{in}>$.
Nevertheless, the considered model shows that during the evolution topology
fluctuations strongly increase. Indeed, in the {\em out} region the "space
density" $N(x)$ turns out to be proportional to the spatial volume $V_x$ of
the configuration manifold $M_x$.  In the given model the volume $V_x$ is
infinite and that of the "space density" but it would not be so if we
consider the real potential in (\ref{eq:wdwinfl}) (or consider next orders of
an approximation procedure) and, therefore, one could expect the value $N(x)$
to be sufficiently large but finite. Then if the initial state corresponds to
a simple universe (\ref{eq:consts}) having $N_{in}(x)=1$ the final state
will be described by the density matrix (\ref{eq:infden}) up to the order of
$1/V$.

\end{document}